# Dislocation-free axial InAs-on-GaAs nanowires on silicon

Daria V. Beznasyuk[1,2], Eric Robin[1,3], Martien Den Hertog[1,2], Julien Claudon[1,4], Moïra Hocevar[1,2]

[1] Université Grenoble-Alpes, F-38000 Grenoble, France

[2] CNRS-Institut Néel, 25 av. des Martyrs, F-38000 Grenoble, France

[3] CEA, INAC-MEM, 17 av. des Martyrs, F-38000 Grenoble, France

[4] CEA, INAC-PHELIQS, 17 av. des Martyrs, F-38000 Grenoble, France

**Abstract**

We report on the growth of axial InAs-on-GaAs nanowire heterostructures on silicon by molecular beam epitaxy using 20 nm diameter Au catalysts. First, the growth parameters of the GaAs nanowire segment were optimized to achieve a pure wurtzite crystal structure. Then, we developed a two-step growth procedure to enhance the yield of vertical InAs-on-GaAs nanowires. We achieved 90% of straight InAs-on-GaAs nanowires by further optimizing the growth parameters. We investigated the composition change at the interface by energy dispersive X-ray spectroscopy and the nanowire crystal structure by transmission electron microscopy. The nominal composition of the InAs segment is found to be $In_xGa_{1-x}As$ with x=0.85 and corresponds to 6% of lattice mismatch with GaAs. Strain mapping performed by the geometrical phase analysis of high-resolution images revealed a dislocation-free $GaAs/In_{0.85}Ga_{0.15}As$ interface. In conclusion, we successfully fabricated highly mismatched heterostructures, confirming the prediction that axial $GaAs/In_{0.85}Ga_{0.15}As$ interfaces are pseudomorphic in nanowires below 40 nm diameter.

**1. Introduction**

Integrating direct bandgap III-V compound semiconductors on silicon is seen as an appealing strategy to realize optical emitters on a chip. Considerable challenges faced by thin film epitaxy of III-V semiconductors on Si due to strain and materials chemistry motivated research towards new fabrication strategies, including wafer bonding [1], selective area growth [2] or nanowire growth [3]. Nanowires have a nanometer-scale cross section which enables the release of strain via lateral elastic relaxation, not only across the nanowire/substrate interface [4] but also across material interfaces in the nanowire axis [5, 6]. Elastic relaxation prevents the formation of dislocations and, in theory, enables combinations of infinitely long semiconductor segments with up to 7 % lattice mismatch (InAs and GaAs for example) in nanowires below 40 nm in diameter [7]. In comparison with two dimensional systems, the critical thickness below which no dislocation appears is only few monolayers for InAs-on-GaAs layers [8]. Yet, experimental data for InAs-on-GaAs nanowires are missing to confirm the prediction of dislocation-free interfaces below critical diameter because the growth is still challenging.

Most axial nanowire heterostructures grow by the vapor-liquid-solid (VLS) mechanism with a gold nanoparticle used as a catalyst [9]. In VLS, the switch from one material to the other is delicate, in particular, when interchanging the group III element (Ga and In) for III-V semiconductor combinations. Those heterostructures are affected by interface grading and kinking [10, 11, 12]. Interface grading is a consequence of both group III elements being soluble in the liquid Au droplet. If the first element is not properly purged from the droplet, it will continue to incorporate in the new segment and form a graded interface, which is detrimental for the control of optical properties. Moreover, kinking of the upper segment happens when the gold nanoparticle is destabilized during the group III element switch. One way to prevent kinking is to use buffer layers of InGaAs between GaAs and InAs [13, 14] but this solution is very similar to having a graded interface as it circumvents from controlling optical confinement for example. Another solution to avoid both interface grading and kinking is to grow Au-free-catalyzed nanowires [15]. However, it still remains unclear whether the formation of thin diameters (and therefore dislocation-free interfaces) is possible by this technique. In a nanowire system using Au catalysts, InAs-on-GaAs and GaAs-on-InAs interface grading can be minimized using growth interrupts, small diameter catalysts and high V/III ratios [10, 16]. Additionally, InAs-on-GaAs interfaces are usually much sharper than GaAs-on-InAs interfaces [17, 18]. This discrepancy was explained by the higher affinity of In with Au [12], making it more difficult to expel from the droplet. Preventing kinking requires to maintain a stable position of the droplet on a flat nanowire top facet during group III element switch. During growth, wurtzite nanowires only nucleate when all edges below the gold droplet are sharp, whereas truncated facets at the liquid solid interface produce zinc-blende nanowires [19]. Therefore, constraining GaAs and InAs segments to the wurtzite crystal phase is one solution to stabilize the droplet on flat facets and increase the vertical yield of InAs-on-GaAs nanowires [20]. The wurtzite crystal structure of arsenide compounds forms for high levels of supersaturation in the droplet [19, 21]. Moreover, high supersaturations produce layer-by-layer growth of the new segment in nanowire heterostructures [22] and minimize the length of truncated edge facets [19], which are suspected to be responsible for the formation of kinks in nanowire heterostructures based on group V elements. Promoting these effects contributes to the droplet mechanical stability and to the growth of straight heterostructures. It was recently proposed that increasing the group III concentration in the catalyst also induces a higher stability in nanowire heterostructures based on group III element [23]. The authors pre-deposited In before the growth of their GaAs nanowires, which led to an increase of the Ga solubility in the Au droplet from 33% to 45% and to an increase of the vertical yield of InAs-on-GaAs nanowires.

Here, we report on the growth of axial InAs-on-GaAs nanowire heterostructures below critical diameter using both the growth temperature and In flux as optimization parameters. We show that by maintaining both a high supersaturation and a high group III concentration in the gold droplet during the whole growth sequence, the yield of straight InAs-on-GaAs nanowires reaches 90%. We measured the composition of the InAs segment, which happens to be $In_xGa_{1-x}As$ with x ≈ 0.85, corresponding to a lattice mismatch of 6% with GaAs. We demonstrate that the interface between GaAs and $In_{0.85}Ga_{0.15}As$ depends on the In flux and the V/III beam equivalent pressure (BEP) ratio and can be as sharp as 5 nm. Finally, strain mapping of the InAs-on-GaAs nanowire revealed a dislocation-free interface.

**2. Experimental details**

*2.1 Growth of wurtzite GaAs nanowires on Si (111) substrates*

We carried out the growth of GaAs nanowires in a molecular beam epitaxy (MBE) reactor. The nanowires were grown via the Au-assisted vapor-liquid-solid (VLS) mechanism on n-type (111)-oriented silicon substrates (figure 1). We chemically treated the substrates prior to their introduction into the MBE reactor. First, a combination of HF and $NH_4F$ was used to remove the native oxide and to create atomically flat terraces on the silicon surface [24]. Then, the substrates were covered with a gold colloid solution and finally rinsed in deionized water. We loaded the samples into the introduction module immediately after their chemical preparation to prevent re-oxidation of the silicon surface. Before reaching the growth chamber, the substrates were degassed at 200 °C in a transfer module until the pressure dropped below $2.0 \times 10^{-8}$ Torr. We grew the GaAs nanowires as follows. The sample was heated to the desired growth temperature with a ramp of 40 °C/min. After one minute of stabilization, we opened the arsenic cracker-cell shutter and, ten seconds later, the gallium shutter. After 30 minutes, we closed the gallium shutter and we finally cooled down the sample under an arsenic flux. To promote the formation of the wurtzite crystalline phase, the gold colloids diameter and the growth temperature were intentionally set to 20 nm and 610°C, respectively [25, 21]. We strategically choose a diameter of 20 nm, since it is below the theoretical critical diameter to prevent the formation of dislocations at the interface between GaAs and InAs and to achieve high group III concentration in the droplet.

In our MBE system, the GaAs deoxidation temperature equals 630 °C. We studied the nanowire morphology and crystal structure for two different V/III BEP ratios at a fixed Ga flux of $2.0 \times 10^{-7}$ Torr corresponding to a growth rate of 0.34 monolayers per second (ML/s) on GaAs. The nanowires were grown at a V/III BEP ratio of 15 and 30 under an As flux of $3.0 \times 10^{-6}$ Torr and $6.0 \times 10^{-6}$ Torr, respectively.

*2.2 Growth of straight InAs-on-GaAs nanowires on Si (111) substrates*

Once we obtained the desired wurtzite GaAs nanowires, we switched to the growth of InAs-on-GaAs nanowires. The growth temperature of InAs nanowires by MBE is significantly lower than that of GaAs nanowires, usually around 410 °C [26]. We introduced a two-step temperature procedure (figure 2), where the GaAs segment grows at high temperature (610°C) and InAs at lower temperatures. The growth time for GaAs was 30 min, including the cooling down process of 5 min.

As growth interruptions are known to improve the abruptness of nanowire heterointerfaces [10, 27], we studied two growth interruption strategies: (a) we interrupted the growth of GaAs during the whole cooling procedure (figure 2a) and (b) we interrupted the growth of GaAs after the cooling procedure (figure 2b) during 2 minutes. In both cases, we interrupted GaAs growth by closing the gallium shutter while maintaining the arsenic flux.

We then grew the InAs nanowire segment by opening the In shutter during 25 minutes. We cooled down the sample after switching off the In shutter while maintaining an As flux down to 400°C. We studied how the nanowire morphology is influenced by the InAs growth temperature and the In flux. First, we grew the InAs segment at a fixed In flux of $3.0\times10^{-7}$ Torr (0.48 ML/s) and a V/III BEP ratio of 20 from 420 °C up to 540 °C. Then, we grew another set of samples at an In flux of $1.2\times10^{-7}$ Torr (0.2 ML/s) and a V/III BEP ratio of 50 from 510 °C up to 610 °C.

*2.3 Characterization of the nanowires*

We characterized the GaAs and InAs nanowire segments length and shape by scanning-electron microscopy (SEM). The yield of straight InAs-on-GaAs segments was also analyzed by SEM. We observed 60-80 nanowires per sample. A CM300 transmission electron-microscope (TEM) equipped with a LaB6 thermionic emitter operated at 300 kV was used for detailed crystalline structure investigations in TEM mode. The chemical composition and the morphology of the nanowires were investigated by energy-dispersive X-ray spectroscopy (EDX) and high angle annular dark field (HAADF) scanning transmission electron microscopy (STEM) on a probe corrected FEI Titan Themis working at 200 kV, equipped with four silicon drift detectors for EDX. Quantification of In L-lines, Ga K-lines, As K-lines and Au L-lines was performed using the Cliff-Lorimer method assuming no X-ray absorption in the investigated samples which is justified considering the relatively high energy of the measured X-ray lines (> 3.3 keV) and the small diameter of the nanowires [28]. The Cliff-Lorimer factors were calibrated on the same equipment, at the same operating conditions, using reference samples of known composition and thickness. Finally, we applied the geometrical phase analysis (GPA) method [29, 30] to our high-resolution STEM images to evaluate the lattice deformation in the nanowire heterostructures.

**3. Results and discussion**

*3.1 Wurtzite GaAs nanowires on a Si substrate*

We first discuss the morphology of GaAs nanowires for different V/III ratios. At a V/III ratio of 15, the nanowires are tapered with a large diameter along the growth axis (figure 1b). The diameters at the basis and at the top are 99±13 nm and 35±3 nm, respectively. We measured an average length of 863±85 nm. When increasing the V/III ratio to 30 (figure 1c), the nanowires show a slight tapering at the basis and a thin and homogeneous diameter along the whole length. All nanowires have an average length and diameter of 1.2±0.1 µm and 25±3 nm, respectively.

During growth, the evolution of the nanowire morphology is governed by the competition between adatom surface diffusion and direct impingement on the droplet (figure 1a) [31]. Generally, low V/III ratios (low As coverage) increase Ga diffusion along the nanowire facets [32], which leads to a combination of lateral and axial growth. Moreover, the catalyst diameter is larger at low V/III ratios for nanowires grown in MBE systems [33], which is in agreement with our observations. In contrast, high V/III ratios (high As coverage) reduce the Ga diffusion along the nanowire facets and thus the catalyst droplet diameter: axial growth due to direct impingement is fostered at the expense of radial growth. In conclusion we succeeded to control the GaAs nanowire morphology by controlling the respective contributions of direct impingement and surface diffusion to the axial growth rate. Additional data in supplementary information (SI) (see figure S1) clearly show the transition between tapered and straight nanowires.

We then study the GaAs nanowires crystal structure by TEM. At a V/III ratio of 15 a major part of the nanowires have the wurtzite crystal structure but exhibit many stacking faults (see figure S2). At a higher V/III ratio of 30, the nanowires grow with a pure wurtzite crystal phase, as observed on the selected area diffraction pattern (figure 1e). Such nanowires have less than four stacking faults per micrometer (see figure S3). To the best of our knowledge, this is the first demonstration of pure wurtzite GaAs nanowires grown on silicon using gold as a catalyst.

*3.2 Influence of the growth parameters on the morphology of InAs-on-GaAs nanowires*

As described earlier, we use a two-step temperature procedure, where the GaAs nanowire stems grow at 610°C and the following InAs segments at a lower temperature. First, we discuss the two different strategies applied to the growth of InAs-on-GaAs nanowires, where the cooling procedure between GaAs and InAs growths takes place under an As flux only (Figure 2a) or during GaAs growth (Figure 2b). After cooling under As only, most InAs segments grow kinked with respect to the GaAs nanowire stem (Figure 2c). Moreover, we notice that the InAs segment has a significantly larger diameter than the GaAs stem (Figure 2e). In contrast, the InAs segments grow straight on top of the GaAs stems when GaAs continues growing during the cooling procedure (Figure 2d). We observe that the InAs-on-GaAs nanowire diameter evolves along the growth axis (Figure 2f): the GaAs nanowire segments have a bottle-shaped morphology, where the base of each nanowire has a larger diameter than the tip. It is an indication of GaAs and/or InAs radial overgrowth. The InAs segment is clearly visible above GaAs with a homogeneous diameter of 37±3 nm along the whole growth axis.

Our observations suggest that the excess of Ga present in the droplet determines whether the InAs segment will kink or not, which is in agreement with the work reported in [23], where the concentration of group III element in the droplet is correlated to its stability. During the cooling step under As, the Ga is purged from the droplet by forming a GaAs nanowire section. In this situation, the droplet size and the contact angle formed with the nanowire top facet (see Figure 1a) decrease, favoring the formation of a GaAs zinc blende section which diameter shrinks below 20 nm [7, 34] (see the cooling neck on figure 1d, as well as the contact angle <90°). At the In shutter opening, a large amount of In dissolves into the depleted droplet both from direct impingement and from adatom surface diffusion. We think that such abrupt volume expansion destabilizes the droplet, which falls on the side, leading to the growth of kinked InAs segments (see figure S4). On the contrary, during the cooling step under Ga and As, the droplet is constantly fed with Ga. Therefore, its size and the contact angle formed with the nanowire top facet remain constant (or decrease to a lesser degree due to the Ga diffusivity drop). Moreover, in Au droplets, the Ga concentration increases exponentially with the decrease of the diameter. Ga solubility reaches 40% for 20 nm diameter droplets [35, 36], which contributes to maintaining a high group III concentration (and supersaturation) in the catalyst. Thus, the GaAs nanowires keep growing in the wurtzite crystal structure, forming sharp edge facets at the liquid-solid interface. We believe that InAs nucleates from a mechanically stable Au droplet on a flat top facet, favoring straight growth. Earlier works on axial heterostructures of energetically non-favorable semiconductor combinations (InAs-on-GaAs [23], Si-on-GaP [22], InP-on-InAs [37]) suggested that a high supersaturation and/or a high group III concentration in the droplet during the upper segment formation favors vertical growth. We will show later that a high group III concentration in the droplet is required not only during the upper segment formation but along the entire interchange sequence to prevent droplet depletion and favor straight growth of InAs-on-GaAs nanowires.

Before optimizing the morphology of the nanowires, we investigate the influence of the growth temperature on the yield of vertical InAs segment (see figure S5). We find a maximum yield of 90% of straight InAs segments at temperatures ranging between 510°C and 570°C for various In fluxes. Below 510 °C, InAs hardly contributes to the axial growth on top of GaAs and mostly forms a thick and irregular shell around the GaAs nanowire (see figure S5a and figure S6 for the EDX measurements). On the contrary, above 570°C, InAs segments grow on top of GaAs. However, the segments kink (see figure S5c and figure S7). At such high temperatures, the In desorption rate increases (yet remains slower than As) resulting in a lower In fraction in the catalyst and correspondingly lower supersaturation. As discussed earlier, low supersaturation favors long truncated edges at the solid/liquid interface resulting in a decrease of droplets stability and a higher probability of kinks. Based on our observations, we set the optimum growth temperature to 540°C for the InAs segment.

We now study whether we can control the nanowire morphology using different In fluxes (Figure 3). As already observed by SEM on figure 2b, an In flux of $3.0 \times 10^{-7}$ Torr leads to a bottle-shape morphology. A 300 nm-long InAs segment grows on top of GaAs (Figure 3a), indicating an InAs axial growth rate of 12 nm/min. Yet, a 40 nm thick InAs shell surrounds the bottom of the GaAs segment (inset of Figure 3a). This shell forms at a lateral rate of 1.6 nm/min. By decreasing the In flux down to $1.2 \times 10^{-7}$ Torr under a constant As flux (thus increasing the V/III ratio from 20 to 50), we observe that the nanowire morphology is more uniform. A 60 nm-long InAs segment grows on top of GaAs (Figure 3b) with an axial growth rate of 2.5 nm/min. The parasitic shell around the GaAs segment seems suppressed (inset Figure 3b). Yet, we measure a shell of 4 nm, which corresponds to a radial growth rate of 0.16 nm/min. The ratio between axial and radial growth rates is twice as large at low In flux compared to high In flux. For InAs segments of similar length obtained at low and high In fluxes, the parasitic shells would be half as thin using a low In flux. Finally, we observe that the InAs segment grown at a high In flux features more stacking faults and crystal phase switches (approximately 15 per 50 nm) along the entire growth axis than the one grown at low In flux (approximately 10 per 50 nm).

In conclusion, using an In flux of $1.2 \times 10^{-7}$ Torr improves the InAs-on-GaAs nanowire morphology and crystalline quality. We can see the similarity with the wurtzite GaAs nanowires: a high As coverage (high V/III ratios) reduces the In surface diffusion and leads to the reduction both of tapering and of crystalline defects. Still, we need further work to fully suppress the crystal phase switching in the InAs segment by fine tuning the In flux and V/III ratio.

*3.3 InAs-on-GaAs nanowire interface sharpness and crystalline quality*

Now that we created straight InAs-on-GaAs nanowires, we first study the interface quality in terms of sharpness. The most direct way is to measure the chemical composition by EDX along the nanowire axis (see paragraph 2.3). Figure 4a shows a HAADF STEM image of our nanowire. The upper segment has a larger diameter and features dark and bright areas which correspond to a change either in material composition or in crystal structure. We studied the average chemical composition along and across the nanowire axis as indicated in Figure 4a. On the axial EDX line profile we clearly see three zones (Figure 4b): the long bottom segment is pure GaAs, the upper segment with a larger diameter is InAs and the droplet is Au. The nominal composition of the InAs segment is a ternary $In_xGa_{1-x}As$ alloy (x varying from 0.9 at the GaAs interface to 0.8 near the top). A pure 5 nm InAs segment is visible below the Au catalyst. The droplet contains 70% of Au and 30% of In. We measure a $GaAs/In_{0.85}Ga_{0.15}As$ interface sharpness length which ranges between 15 and 30 nm (measured on several nanowires). The EDX line profiles taken perpendicular to the growth axis of the same InAs-on-GaAs nanowire are presented in Figure 4 c,d. We find that a few nm-thick $In_{0.5}Ga_{0.95}As$ shell surrounds the GaAs segment (Figure 4c). Finally, we observed that the samples grown with an In flux of $3\times10^{-7}$ Torr (Figure 3a) feature sharper $GaAs/In_{0.85}Ga_{0.15}As$ interfaces down to 5 nm (see figure S8).

The interface length is significantly reduced when using higher In flux, approaching the values obtained for self-catalysed InAs-on-GaAs nanowires [15]. By introducing In in the gold droplet, the Ga solubility decreases. The faster In enters the Au (higher In flux), the faster Ga is depleted and the sharper is the interface. However, the question remains why the expected InAs segment is in fact a ternary alloy and not pure InAs, as seen in the EDX line profile in Figure 4b (and S8b)? The $In_{0.85}Ga_{0.15}As$ segment has a constant composition in the radial direction, confirming that Ga adatoms are incorporated in the volume (Figure 4c). One explanation can be found in our two-step growth procedure where GaAs continues to grow during the temperature decrease down to 540 °C. Ga adatoms do not desorb from the surface during the group III interchange at 540°C [38], creating a reservoir of available Ga adatoms. Therefore, residual Ga diffuses to the Au droplet and contributes to the formation of the $In_xGa_{1-x}As$ segment by the VLS mechanism. We interpret the observation of a decrease in In composition along the $In_xGa_{1-x}As$ by the presence of a competing mechanism between In and Ga adatom diffusion from the surface. We suggest that another source of Ga in the $In_xGa_{1-x}As$ segment could be the diffusion of Ga from the bottom GaAs segment across the $GaAs/In_xGa_{1-x}As$ interface, as observed in Stransky-Krastanov InAs quantum dots grown on GaAs [39]. Growing InAs segments during different times, at different In fluxes and using different cooling down rates would provide useful data to conclude on the diffusivity competition between In and Ga and its dependence on temperature, as well as on the diffusivity of Ga across GaAs/InAs interfaces. The formation of a thin shell around the bottom segment is usual in axial nanowire heterostructures and can be suppressed via fine tuning of the growth parameters. After optimization, the shell is still $In_{0.5}Ga_{0.95}As$, which further confirms the presence of mobile Ga adatoms and competing mechanisms between In and Ga. Finally, the presence of the pure InAs cooling neck is probably related to In segregation [40] or rapid crystallization from the AuIn alloy.

We are now interested in the crystal structure of the nanowires, especially in the structural quality of the $GaAs/In_{0.85}Ga_{0.15}As$ interface. Figure 5a shows a bright field high-resolution TEM image of a structure. The major part of the nanowire has the wurtzite crystalline structure (see figure S9 for the diffraction patterns). The GaAs segment is defect free, as already seen in Figure 1d, yet the $In_{0.85}Ga_{0.15}As$ segment features randomly distributed zinc-blende inclusions. Interestingly, we clearly see an expansion of the $In_{0.85}Ga_{0.15}As$ diameter in Figures 5 a and b. The GaAs and $In_{0.85}Ga_{0.15}As$ segments measure 32.6 nm and 38 nm in diameter, respectively. This diameter expansion was explained earlier by the variation of the catalyst composition and thus of the contact angle [41, 42, 43]. Here, $In_{0.85}Ga_{0.15}As$ has a lattice parameter ~6% larger than GaAs: a large strain builds across the $GaAs/In_{0.85}Ga_{0.15}As$ interface and has to be released to minimize the energy of the system. We calculate that for a certain number of planes in a GaAs segment of 33 nm in diameter (a = 0.39845 nm in wurtzite GaAs), the diameter expansion is 2.36 nm for a fully relaxed InAs segment containing the same number of planes (a = 0.4269 nm). We therefore believe that strain relaxation at the interface is partly responsible for the diameter variation.

The careful visual inspection of high-resolution STEM images (Figure 5c, see figure S9c for the whole interfacial region) did not reveal any misfit dislocations in the crystal at the interface between the GaAs and In$_{0.85}$Ga$_{0.15}$As segments, though the crystal phase switches from WZ to ZB in the transition region. We found this effect for all investigated nanowire heterostructures both for low and high In fluxes. We speculate it could be an additional source to release strain for the system.

Finally, we applied the GPA method to the high resolution HAADF image in Figure 5b using the (0002) reflection, to map the local interplanar spacing at the GaAs/In$_{0.85}$Ga$_{0.15}$As interface (Figure 5d). The orange top segment is In$_{0.85}$Ga$_{0.15}$As and the green bottom segment is GaAs. The interface does not present any sign of abrupt color variation, which indicates the absence of misfit dislocations [44, 45].

## 4. Conclusion

In conclusion, we have successfully fabricated dislocations free axial InAs-on-GaAs nanowire heterostructures on silicon with a yield up to 90%. By a careful optimization of the growth parameters, we obtained for the first time pure GaAs wurtzite nanowires with Au as a catalyst on Si by MBE. We then proposed a two-step growth procedure to stabilize the Au catalyst during the group III element interchange by constantly maintaining a high supersaturation. By decreasing the In flux and increasing the V/III ratio, we reduced the formation of a radial shell around the GaAs stem as well as the density of crystal defects. We performed EDX, HRTEM and GPA and demonstrated that our GaAs/In$_{0.85}$Ga$_{0.15}$As interfaces are free of dislocations despite a ~6% lattice mismatch. Such interface control of high lattice mismatched materials is crucial for the design of future optical emitters on Si substrates.

**Acknowledgments**

This work was supported by LANEF PhD Program and by the University of Grenoble-Alpes through the AGIR program. The authors are grateful to Yann Genuist and Didier Boilot for their MBE technical support.

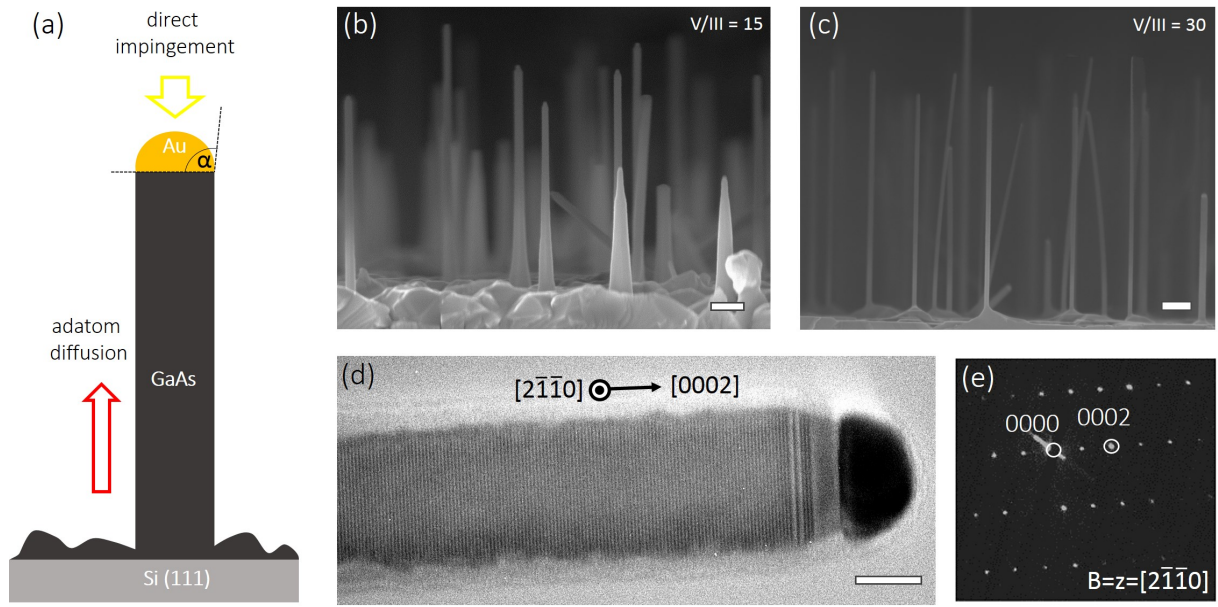

**Figure 1. Au-GaAs wurtzite nanowires on Si substrates.** a) Schematic model of VLS GaAs nanowire growth on silicon. Two major contributions are shown: direct impingement into the catalyst and adatom diffusion along the nanowire facets. The contact angle is defined as α. SEM images (side view, scale bar 200 nm) of GaAs nanowires grown with (b) a V/III BEP ratio of 15 and (c) a V/III BEP ratio of 30. d) Bright-field TEM image (scale bar 10 nm) of an individual GaAs nanowire from (c). The wire has a wurtzite crystal structure as shown on the electron diffraction pattern (e). The region directly below the gold particle features a "cooling neck" with a zing-blende crystal phase.

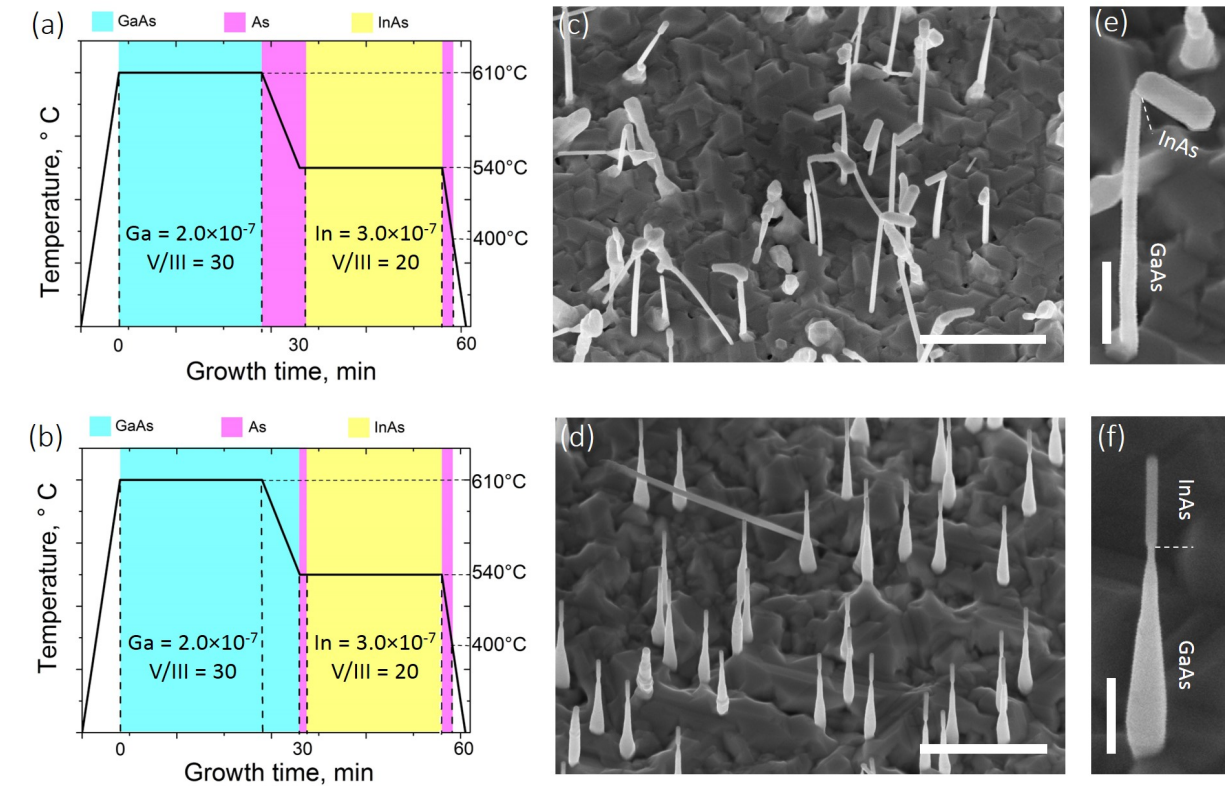

**Figure 2. Comparison of two different cooling strategies applied to the growth of InAs-on-GaAs nanowires.** InAs-on-GaAs nanowire growth sequence with intermediate cooling down (a) under As flux and (b) under continuous growth of GaAs. (c) and (d) SEM images (30° tilt, scale bar 1 µm) of InAs-on-GaAs nanowires corresponding to the growth sequences (a) and (b), respectively. (e) and (f) Magnified SEM images (30° tilt, scale bar 200 nm) of a single InAs-on-GaAs nanowire from image (c) and (d), respectively.

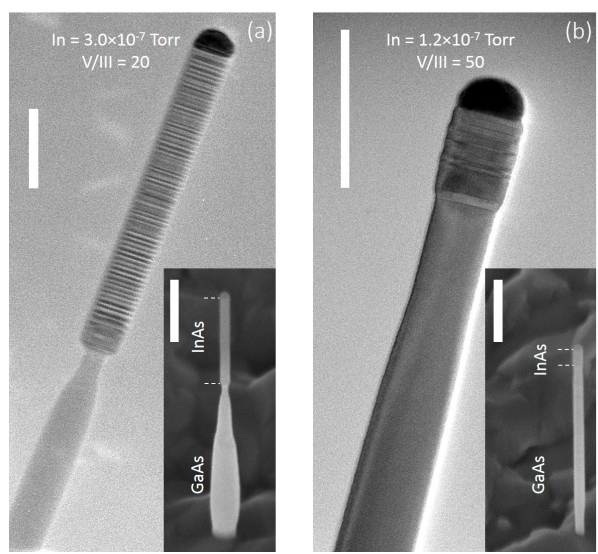

**Figure 3. Morphology of InAs-on-GaAs nanowires grown at 540 °C for two different In fluxes.** a) Bright-field TEM image (scale bar 100 nm) of an InAs-on-GaAs nanowire. Inset: SEM image of an InAs-on-GaAs from the same sample. The scale bar is 200 nm, 30° tilt. The bottle shape of GaAs segment indicates radial overgrowth. b) Bright-field TEM image (scale bar 100 nm) of an InAs-on-GaAs nanowire grown with lower In flux compared to (a). Inset: SEM image of an InAs-on-GaAs from the same sample. The scale bar is 200 nm, 30° tilt. Note that parasitic 2D growth on the substrate is 380 nm for (a) and 330 nm for (b): it is the reason why the GaAs segment looks longer in (b) than in (a).

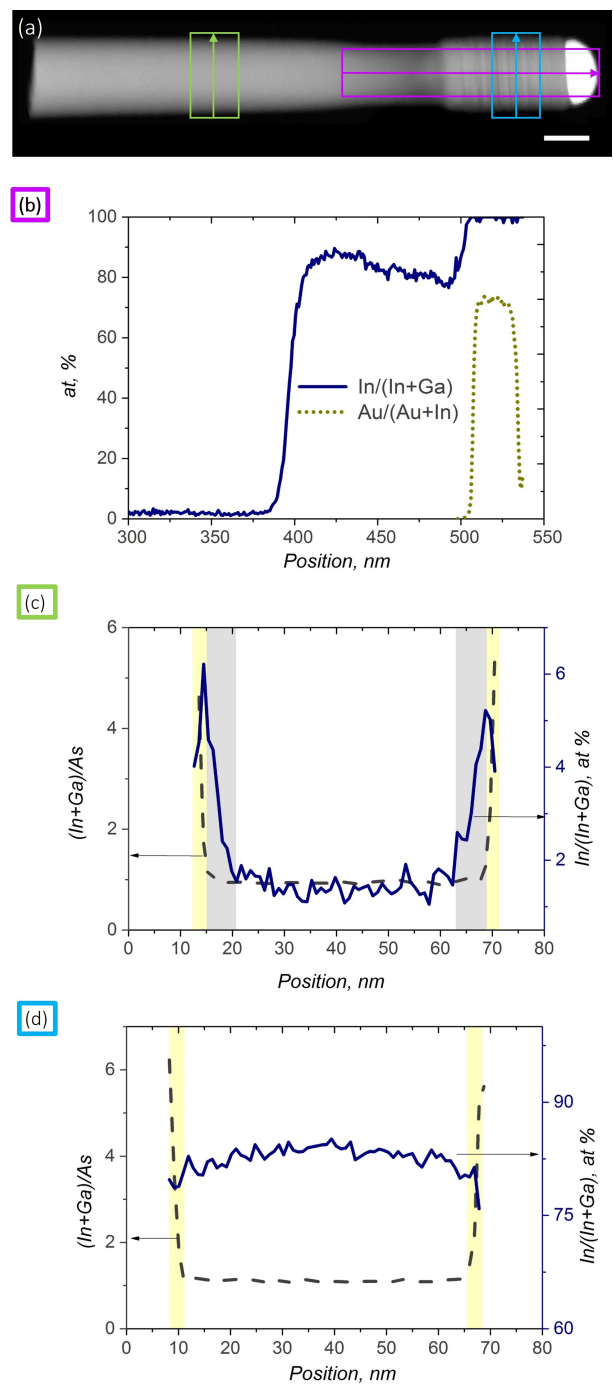

**Figure 4. Investigation of the nanowire chemical composition.** a) HAADF-STEM image (scale bar 30 nm) of an InAs-on-GaAs nanowire grown at 540 °C with In = 1.2×10$^{-7}$ Torr and V/III ratio of 50. b) EDX line profile of In and Au composition along the growth axis of the nanowire from image (a). c),d) EDX line profiles of In and As across the nanowire diameter from image (a). A thin In$_{0.5}$Ga$_{0.95}$As shell is highlighted in gray and a layer of oxide in yellow.

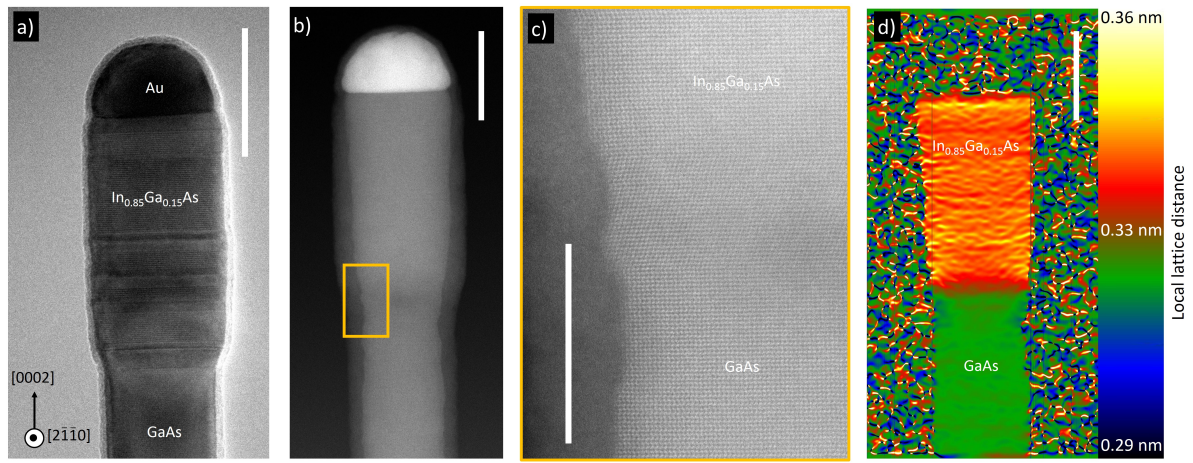

**Figure 5. Investigation of the nanowire crystal structure.** a) Bright-field TEM image in the <2-1-10> viewing direction (scale bar 40 nm) of an InAs-on-GaAs nanowire grown at 540 °C with In = $1.2\times10^{-7}$ Torr and V/III ratio of 50. b) High resolution HAADF-STEM image (scale bar 30 nm) of a nanowire from the same sample. c) Close up view (scale bar 10 nm) of the nanowire interface from (b). d) Local lattice distance (scale bar 30 nm) obtained by GPA. We applied GPA to the 0002 spot in the Fast Fourier Transform (FFT) of image (b). The orange region corresponds to the InAs segment and the green one to GaAs. Continuous change in color at the GaAs/In$_{0.85}$Ga$_{0.15}$As interface indicates dislocation-free InAs-on-GaAs nanowire.

# References

[1] Tanabe, K., Watanabe, K. and Arakawa, Y., *"III-V/Si hybrid photonic devices by direct fusion bonding."*, Scientific reports 2, 349, 2012.

[2] Li, R.R., Dapkus, P.D., Thompson, M.E., Jeong, W.G., Harrison, C., Chaikin, P.M., Register, R.A. and Adamson, D.H., *"Dense arrays of ordered GaAs nanostructures by selective area growth on substrates patterned by block copolymer lithography.",* Applied Physics Letters 76(13), 1689-1691, 2000.

[3] Plissard, S., Dick, K. A., Larrieu, G., Godey, S., Addad, A., Wallart, X., and Caroff, P., *"Gold-free growth of GaAs nanowires on silicon: arrays and polytypism",* Nanotechnology 21(38), 385602, 2010.

[4] Cirlin, G.E., Dubrovskii, V.G., Soshnikov, I.P., Sibirev, N.V., Samsonenko, Y.B., Bouravleuv, A.D., Harmand, J.C. and Glas, F., "Critical diameters and temperature domains for MBE growth of III–V nanowires on lattice mismatched substrates," *physica status solidi (RRL)-Rapid Research Letters, 3(4), 112-114,* 2009.

[5] Ercolani, D., Rossi, F., Li, A., Roddaro, S., Grillo, V., Salviati, G., Beltram, F. and Sorba, L., "2009. InAs/InSb nanowire heterostructures grown by chemical beam epitaxy," *Nanotechnology, 20(50), 505605,* 2009.

[6] Kaganer, V.M. and Belov, A.Y., "Strain and x-ray diffraction from axial nanowire heterostructures," *Physical Review B, 85(12):125402,* 2012.

[7] Glas, F., *"Critical dimensions for the plastic relaxation of strained axial heterostructures in free-standing nanowires",* Physical Review B 74(12), 121302., 2006.

[8] Grandjean, N., and Massies, J., "Epitaxial growth of highly strained $In_xGa_{1-x}As$ on"GaAs (001): the role of surface diffusion length," *Journal of crystal growth 134(1-2), 51-62,* 1993.

[9] Lauhon, L.J., Gudiksen, M.S. and Lieber, C.M., "Semiconductor nanowire heterostructures.," *Philosophical Transactions of the Royal Society of London A: Mathematical, Physical and Engineering Sciences, 362(1819), 1247-1260.,* 2004.

[10] Dick, K.A., Bolinsson, J., Borg, B.M. and Johansson, J., "Controlling the abruptness of axial heterojunctions in III–V nanowires: beyond the reservoir effect," *Nano letters, 12(6), 3200-3206.,* 2012.

[11] Paladugu, M., Zou, J., Guo, Y.N., Auchterlonie, G.J., Joyce, H.J., Gao, Q., Hoe Tan, H., Jagadish, C. and Kim, Y., "Novel growth phenomena observed in axial InAs/GaAs nanowire heterostructures," *Small, 3(11), pp.1873-1877.,* 2007.

[12] Paladugu, M., Zou, J., Guo, Y.N., Zhang, X., Kim, Y., Joyce, H.J., Gao, Q., Tan, H.H. and Jagadish, C., "Nature of heterointerfaces in GaAs/InAs and InAs/GaAs axial nanowire heterostructures," *Applied Physics Letters, 93(10), 101911,* 2008.

[13] Xin, Y., Xia, Z., Jun-Shuai, L., Xiao-Long, L., Xiao-Min, R. and Yong-Qing, H., "Growth and characterization of straight InAs/GaAs nanowire heterostructures on Si substrate," *Chinese Physics B, 22(7), 076102,* 2013.

[14] Huang, H., Ren, X., Ye, X., Guo, J., Wang, Q., Zhang, X., Cai, S. and Huang, Y., "Control of the crystal structure of InAs nanowires by tuning contributions of adatom diffusion," *Nanotechnology, 21(47), 475602,* 2010.

[15] Scarpellini, D., Somaschini, C., Fedorov, A., Bietti, S., Frigeri, C., Grillo, V., Esposito, L., Salvalaglio, M., Marzegalli, A., Montalenti, F. and Bonera, E., "InAs/GaAs sharply defined axial heterostructures in self-assisted nanowires," *Nano letters, 15(6), 3677-3683.,* 2015.

[16] Priante, G., Patriarche, G., Oehler, F., Glas, F. and Harmand, J.C., "Abrupt GaP/GaAs interfaces in self-catalyzed nanowires," *Nano letters, 15(9), 6036-6041,* 2015.

[17] Krogstrup, P., Yamasaki, J., Sørensen, C.B., Johnson, E., Wagner, J.B., Pennington, R., Aagesen, M., Tanaka, N. and Nygård, J., "Junctions in axial III– V heterostructure nanowires obtained via an interchange of group III elements," *Nano letters, 9(11), 3689-3693,* 2009.

[18] Venkatesan, S., Madsen, M.H., Schmid, H., Krogstrup, P., Johnson, E. and Scheu, C., "Direct observation of interface and nanoscale compositional modulation in ternary III-As heterostructure nanowires." *Applied Physics Letters, 103(6), 063106.,* 2013.

[19] Jacobsson, D., Panciera, F., Tersoff, J., Reuter, M.C., Lehmann, S., Hofmann, S., Dick, K.A. and Ross, F.M., "Interface dynamics and crystal phase switching in GaAs nanowires," *Nature, 531(7594), 317-322.,* 2016.

[20] Messing, M.E., Wong-Leung, J., Zanolli, Z., Joyce, H.J., Tan, H.H., Gao, Q., Wallenberg, L.R., Johansson, J. and Jagadish, C., "Growth of straight InAs-on-GaAs nanowire heterostructures," *Nano letters, 11(9), 3899-3905.,* 2011.

[21] Glas, F., Harmand, J.C. and Patriarche, G., "Why does wurtzite form in nanowires of III-V zinc blende semiconductors?," *Physical review letters, 99(14), 146101.,* 2007.

[22] Hocevar, M., Immink, G., Verheijen, M., Akopian, N., Zwiller, V., Kouwenhoven, L. and Bakkers, E., "Growth and optical properties of axial hybrid III-V/Si nanowires," *Nature Commun. 3 737,* 2012.

[23] Zannier, V., Ercolani, D., Gomes, U.P., David, J., Gemmi, M., Dubrovskii, V.G. and Sorba, L., "Catalyst composition tuning: the key for the growth of straight axial nanowire heterostructures with group III interchange," *Nano Letters, 16(11), 7183-7190,* 2016.

[24] Higashi, G.S., Becker, R.S., Chabal, Y.J. and Becker, A.J., "Comparison of Si (111) surfaces prepared using aqueous solutions of NH4F versus HF," *Applied Physics Letters, 58(15), 1656-1658,* 1991.

[25] Tchernycheva, M., Harmand, J.C., Patriarche, G., Travers, L. and Cirlin, G.E., "Temperature conditions for GaAs nanowire formation by Au-assisted molecular beam epitaxy," *Nanotechnology, 17(16), 4025,* 2006.

[26] Tchernycheva, M., Travers, L., Patriarche, G., Glas, F., Harmand, J.C., Cirlin, G.E. and Dubrovskii, V.G., "Au-assisted molecular beam epitaxy of InAs nanowires: Growth and theoretical analysis," *Journal of Applied Physics, 102(9), 094313.,* 2007.

[27] Dubrovskii, V.G. and Sibirev, N.V. , "Factors Influencing the Interfacial Abruptness in Axial III–V Nanowire Heterostructures," *Crystal Growth & Design, 16(4), 2019-2023,* 2016.

[28] Cliff, G. and Lorimer, G. , "The quantitative analysis of thin specimens," *Journal of Microscopy, 103(2), 203-207,* 1975.

[29] Hÿtch, M.J., Snoeck, E. and Kilaas, R., "Quantitative measurement of displacement and strain fields from HREM micrographs," *Ultramicroscopy, 74(3), 131-146,* 1998.

[30] Rouviere, J.L. and Sarigiannidou, E., "Theoretical discussions on the geometrical phase analysis," *Ultramicroscopy, 106(1), 1-17,* 2005.

[31] Dubrovskii, V.G. and Sibirev, N.V., "General form of the dependences of nanowire growth rate on the nanowire radius," *Journal of crystal growth, 304(2), 504-513.,* 2007.

[32] Nomura, Y., Morishita, Y., Goto, S., Katayama, Y. and Isu, T., "Surface diffusion length of Ga adatoms on (111) B surfaces during molecular beam epitaxy," *Applied physics letters, 64(9), pp.1123-1125.,* 1994.

[33] Paek, J.H., Nishiwaki, T., Yamaguchi, M. and Sawaki, N., "MBE-VLS growth of GaAs nanowires on (111) Si substrate," *physica status solidi (c), 5(9), 2740-2742.,* 2008.

[34] Persson, A.I., Larsson, M.W., Stenström, S., Ohlsson, B.J., Samuelson, L. and Wallenberg, L.R., "Solid-phase diffusion mechanism for GaAs nanowire growth." *Nature Materials, 3(10), 677-681,* 2004.

[35] Zhou, C., Zheng, K., Lu, Z., Zhang, Z., Liao, Z., Chen, P., Lu, W. and Zou, J., "Quality control of GaAs nanowire structures by limiting as flux in molecular beam epitaxy," *The Journal of Physical Chemistry C, 119(35), 20721-20727,* 2015.

[36] Han, N., Wang, F., Hou, J.J., Yip, S., Lin, H., Fang, M., Xiu, F., Shi, X., Hung, T. and Ho, J.C., "2012. Manipulated growth of GaAs nanowires: controllable crystal quality and growth orientations via a supersaturation-controlled engineering process," *Crystal Growth & Design, 12(12), 6243-6249,* 2012.

[37] Svensson, S.F., Jeppesen, S., Thelander, C., Samuelson, L., Linke, H. and Dick, K.A., , "Control and understanding of kink formation in InAs–InP heterostructure nanowires," *Nanotechnology, 24(34), 345601,* 2013.

[38] T Ohachi, J.M Feng, K Asai, "Arsenic pressure dependence of Ga desorption from MBE high index GaAs substrates," *Journal of Crystal Growth ,* Vols. 211 (1–4), 405-410, 2000.

[39] Joyce, P.B., Krzyzewski, T.J., Bell, G.R., Joyce, B.A. and Jones, T.S., "Composition of InAs quantum dots on GaAs (001): Direct evidence for (In, Ga) As alloying," *Physical Review B, 58(24), R15981,* 1998.

[40] Muraki, K., Fukatsu, S., Shiraki, Y. and Ito, R., "Surface segregation of In atoms during molecular beam epitaxy and its influence on the energy levels in InGaAs/GaAs quantum wells," *Applied Physics Letters, 61(5), 557-559,* 1992.

[41] Borg, B.M. and Wernersson, L.E., "Synthesis and properties of antimonide nanowires," *Nanotechnology, 24(20), 202001,* 2013.

[42] Jeppsson, M., Dick, K.A., Wagner, J.B., Caroff, P., Deppert, K., Samuelson, L. and Wernersson, L.E., "GaAs/GaSb nanowire heterostructures grown by MOVPE," *Journal of Crystal Growth, 310(18), 4115-4121,* 2008.

[43] Caroff, P., Wagner, J.B., Dick, K.A., Nilsson, H.A., Jeppsson, M., Deppert, K., Samuelson, L., Wallenberg, L. and Wernersson, L.E., "High-Quality InAs/InSb Nanowire Heterostructures Grown by Metal–Organic Vapor-Phase Epitaxy," *Small, 4(7), 878-882,* 2008.

[44] Frigeri, C., Scarpellini, D., Fedorov, A., Bietti, S., Somaschini, C., Grillo, V., Esposito, L., Salvalaglio, M., Marzegalli, A., Montalenti, F. and Sanguinetti, S. , "Structure, interface abruptness and strain relaxation in self-assisted grown InAs/GaAs nanowires," *Applied Surface Science, 395, 29-36,* 2017.

[45] De la Mata, M., Magén, C., Caroff, P. and Arbiol, J., "Atomic scale strain relaxation in axial semiconductor III–V nanowire heterostructures," *Nano letters, 14(11), 6614-6620,* 2014.